\documentclass[floatfix,aps,amsmath,prl,twocolumn,showpacs,footinbib,reprint]{revtex4-1}
\usepackage{hyperref,url}
\usepackage[hyphenbreaks]{breakurl}
\usepackage{graphicx,graphics,amsfonts,amsmath,mathrsfs,amssymb,bm,psfrag,natbib,dcolumn,textcase}
\usepackage{enumerate}
\usepackage{sidecap}
\allowdisplaybreaks

\newcommand{\be}{\begin{equation}}
\newcommand{\ee}{\end{equation}}

\newcommand{\bs}{\begin{subequations}}
\newcommand{\es}{\end{subequations}}

\begin{document}

\title{Systematic Parameter Errors in Inspiraling Neutron Star Binaries}
\author{Marc Favata}
\email{marc.favata@montclair.edu}
\affiliation{Mathematical Sciences Department, Montclair State University, 1 Normal Avenue, Montclair, NJ 07043, USA}
\affiliation{Theoretical Astrophysics, 350-17, California Institute of Technology, Pasadena, CA 91125, USA}
\affiliation{Department of Physics, University of Wisconsin\mbox{--}Milwaukee, Milwaukee, WI 53201, USA}
\date{Submitted 30 October 2013}
\begin{abstract}
The coalescence of two neutron stars is an important gravitational wave source for LIGO and other detectors.  Numerous studies have considered the \emph{precision} with which binary parameters (masses, spins, Love numbers) can be measured. Here I consider the \emph{accuracy} with which these parameters can be determined in the presence of systematic errors due to waveform approximations. These approximations include truncation of the post-Newtonian (PN) series and neglect of neutron star (NS) spin, tidal deformation, or orbital eccentricity. All of these effects can yield systematic errors that exceed statistical errors for plausible parameter values. In particular, neglecting spin, eccentricity, or high-order PN terms causes a significant bias in the NS Love number.  Tidal effects will not be measurable with PN inspiral waveforms if these systematic errors are not controlled.
\end{abstract}
\maketitle
\setlength\abovedisplayskip{0pt}
\setlength\belowdisplayskip{0pt}
\emph{Introduction}.---One of the key goals of LIGO, Virgo, and other ground-based gravitational-wave (GW) detectors \cite{aLIGOref-harryCQG2010,*Virgo-ref-JINST2012,*Kagraref-somiyaCQG2012,*ETref-CQG2010} is to measure the intrinsic parameters of coalescing neutron star (NS) binaries. The most interesting of these parameters $\theta_a$ are the individual masses $m_i$, the spin angular momenta ${\bm S}_i=\chi_i m_i^2 \hat{{\bm s}}_i$, the orbital eccentricity $e_0$ (at a reference frequency), and the tidal deformability parameters $\lambda_i$ (which depend on the NS masses and equation of state). (Here $i=1,2$ labels the two bodies, $\chi_i$ are the dimensionless spin parameters, $\hat{{\bm s}}_i$ are unit vectors in the spin directions, and $G=c=1$.) Understanding how well we can extract these parameters from a noisy GW signal is especially important as we rapidly approach the operational phase of second-generation detectors.

Many studies have examined how \emph{precisely} LIGO and other detectors will be able to measure the source parameters (e.g., \cite{flanagancutler,*poisson-will-2PNparameterestimate,LIGO-PEpaper2013}).
The achievable precision is ultimately determined by the signal-to-noise ratio (${\rm SNR}$); the corresponding statistical errors due to random noise in the detector scale like $\delta \theta_a \propto {\rm SNR}^{-1} [1 + O({\rm SNR})]$ \cite{vallisneri-fisherabuse-PRD2008}.
A separate but equally important issue is how \emph{accurately} we can measure parameters, i.e., quantifying the systematic bias between the true and best-fit parameters, $\Delta \theta_a\equiv \theta_a^{\rm T} - \theta_a^{\rm best\,fit}$. This systematic error (which is SNR independent) arises from not fully understanding the detector (i.e., calibration error \cite{vitale-etal-LIGOcalibrationerrorPRD2012}) or from deviations between the \emph{true} GW signal $h_{\rm T}(\theta_a^{\rm T})$ and an \emph{approximate} template waveform $h_{\rm AP}(\theta_a^{\rm best\,fit})$ used in the data analysis \cite{canitrot-systematicPE-PRD2001,*brady-fairhurst-systematic-CQG2008,*ajith-systematicerrors-CQG2010,*tyson-etal-systematicPE-2012,*ohme-etal-systematicerrors-PCA-2013,LIGO-PEpaper2013,buonanno-iyer-oshsner-pan-sathya-templatecomparison-PRD2009}. Here I investigate the latter and quantify the systematic bias induced in the intrinsic source parameters due to several potential sources of waveform template errors.
The objective is to understand the costs in parameter inaccuracies if certain physical effects in binary neutron star (BNS) models are neglected.

This study focuses on the inspiral phase of BNS coalescence [which is well described by post-Newtonian (PN) waveforms \cite{blanchetLRR} up to $\sim 1000$ Hz \cite{bernuzzi-nagar-brugmann-PRD2012-tidalNS}]. Since the phasing of the GW signal is significantly more important for parameter estimation than its amplitude, I focus on the following sources of waveform phasing errors:

(a) \emph{High PN-order terms}: For two orbiting point particles with arbitrary masses, the waveform phasing is known completely to 3.5PN order [i.e., order $(v^2)^{3.5}$ beyond the leading-order contribution]. However, unknown terms at 4PN and higher orders may significantly affect parameter accuracy. I use results from analytic BH perturbation theory \cite{saski-tagoshi-LRR} (in which the phasing is known to 22PN order \cite{fujita-14PNEdot,*fujita-22PN}) as an approximate proxy for these unknown terms. Neglecting 4PN and higher-order terms causes systematic errors that exceed statistical ones.

(b) \emph{Spin}: The predicted maximum NS spin is $\chi \sim 0.77$ \cite{cook-shapiro-teukolsky-NRspins-ApJ1994}. NSs are observed to have dimensionless spins $\chi \sim 10^{-4} \mbox{--} 0.3$ \footnote{The pulsar spin distribution is bimodal with peaks near $\chi \sim 10^{-3}$ and $0.1$; observed double NS binaries have spins in the range $\chi \sim 10^{-4} \mbox{--} 0.02$.}. While spin in BNSs is often neglected when modeling their GWs, I show below that even relatively small spins ($\chi \gtrsim 0.003$) can cause non-negligible parameter estimation bias.

(c) \emph{Eccentricity}: While GW emission reduces eccentricity \cite{peters}, compact objects formed in dense stellar environments could have non-negligible eccentricity in the LIGO band \cite{antonini-peretsApJ2012,*antonini-murray-mikkola-ApJ2014,*seto-kozaiPRL2013,*samsin-macleod-ramirez2013,*grindlay-etal2006NatPh,*[{Table 4 of \, }][{ \, indicates that the peak in the BNS eccentricity distribution at 10 Hz occurs at $e_0 \approx 10^{-4}$, with up to $\sim 2\%$ of systems having $e_0>10^{-2}$.}]kowalska-etal-eccentricity-distribution-AA2011}. Neglecting small ($e_0\lesssim 0.02$) eccentricities will not constrain the \emph{detection} of GWs \cite{huerta-brown2013PRD}, but I show that eccentricities $e_0 \gtrsim {\rm few} \times 10^{-3}$ will affect parameter estimation.

(d) \emph{Tides}: When BNS orbital separations are small, each star is tidally distorted by its companion. The resulting change in the gravitational potential modifies the orbital motion and corresponding GW signal. When the orbital period is much longer than the period of stellar oscillation modes that couple to tides (the adiabatic approximation \cite{flanagan-hinderer-lovenumPRD2008}), the effect on the GW phasing can be parameterized by the dimensionless tidal deformation parameters $\hat{\lambda}_i = \lambda_i/m_i^5$. For each star $\lambda\equiv \frac{2}{3} k_2 R^5$ is defined by ${\mathcal I}_{jk} = - \lambda {\mathcal E}_{jk}$, where $k_2$ is the quadrupolar Love number, $R$ is the NS radius, ${\mathcal I}_{jk}$ is the mass quadrupole moment, and ${\mathcal E}_{jk}\equiv R_{tjtk}$ is the quadrupolar electric-type tidal field (i.e., the indicated Riemann tensor components of the companion's spacetime evaluated at the star's center) \cite{[{${\mathcal I}_{jk}$ and ${\mathcal E}_{jk}$ are defined in terms of the $O(r^{-3})$ and $O(r^2)$ coefficients, respectively, of a multipole decomposition of the $g_{tt}$ component of the metric, where $r$ is the distance from the center of mass of one of the stars; see, e.g., Eq.~(D2) of \,}][{}]favatawmm}.
As the NS equation of state is uncertain at high densities, determining the precision with which $\hat{\lambda}_i$ can be constrained has been the focus of several studies (e.g., \cite{hinderer-etal-lovenum-PRD2010,flanagan-hinderer-lovenumPRD2008,*Baiotti-damour-giacomazzo-nagar-rezzolla-tidal-PRL2010,*damour-nagar-villain-PRD2012-measurelovenum,*lackey-shibata-etal-PRD2012-NSBH-eos,*delPozzo-etal-tidal-PRL2013,*hotokezaka-kyutoku-shibata-NRtidaleffects-PRD2013,*read-etal-mattereffectsNSNS-2013,*radice-rezzollaMNRAS2013}). I show that neglecting tidal interactions will cause a small parameter bias. Furthermore, the determination of $\hat{\lambda}_i$ is itself subject to large systematic bias if any of items (a)\mbox{--}(c) above are neglected. Tidal interactions affect the waveform phasing at 5PN and higher orders. Large parameter biases in $\hat{\lambda}_i$ arise when neglected PN terms occur at lower or comparable orders: point-particle terms are not completely known at 4PN order, while spin and eccentricity modify the phasing beginning at 1.5 and 0PN orders, respectively. The remainder of this Letter discusses the waveform models in more detail, describes the formalism for computing statistical and systematic errors, and presents the corresponding results.

\emph{Waveform model}.---The GW signal is modeled using the restricted stationary phase approximation (SPA) in which the Fourier transform (denoted by tildes) of the \emph{true} GW signal is expressed as a function of GW frequency $f$ via $\tilde{h}_{\rm T}(f) = {\mathcal A} f^{-7/6} e^{i \Psi_{\rm T}}(f)$. Here ${\mathcal A} \propto {\rm SNR}$ is a constant depending on the source distance, masses, and orientation angles and does not affect our analysis. The phase is a sum of several possible contributions:
\begin{multline}
\label{eq:Psiterms}
\Psi_{\rm T}(f) = \phi_c + 2\pi f t_c + \frac{3}{128 \eta v^5} \Big( \Delta \Psi_{\rm 3.5PN}^{\rm pp}
\\
+ \Delta \Psi_{\rm 3PN}^{\rm spin} + \Delta \Psi_{\rm 2PN}^{\rm ecc.} + \Delta \Psi_{\rm 6PN}^{\rm tidal}  + \Delta \Psi_{\rm 6PN}^{\rm tm} \Big),
\end{multline}
where $\eta=m_1 m_2/M^2$ is the reduced mass ratio, $M=m_1 + m_2$, $t_c$ and $\phi_c$ are the coalescence time and phase, and $v\equiv (\pi M f)^{1/3}$ is the PN orbital velocity parameter.

The standard 3.5PN point-particle contribution is $\Delta \Psi_{\rm 3.5PN}^{\rm pp} = 1 + \sum_{n=2}^{7} c^{\rm pp}_n(\eta) v^n$, where the $c^{\rm pp}_n(\eta)$ can be found in Eq.~(3.18) of \cite{buonanno-iyer-oshsner-pan-sathya-templatecomparison-PRD2009} and the 2.5 and 3PN coefficients also depend on $\ln v$.

Spin effects to 3PN order are encapsulated in the term
\begin{multline}
\Delta \Psi_{\rm 3PN}^{\rm spin} = 4 \beta_{1.5} v^3 -10 \sigma v^4 + v^5 \ln v^3 \left[ \frac{40}{9} \beta_{2.5}
\right. \\ \left.
- \beta_{1.5} \left( \frac{3715}{189} + \frac{220}{9}\eta \right) \right] + v^6 \left( 160 \pi \beta_{1.5} +20 \beta_{3.0} \right).
\end{multline}
Here $\beta_{1.5}$ is the 1.5PN spin-orbit term \cite{poisson-BHpertIV-slowrotPRD1993,kidder-spineffects};
$\sigma = \sigma_{S_1 S_2} + \sigma_{\rm QM}+ \sigma_{\rm self\,spin}$ is the 2PN spin-spin term which combines three effects \cite{gergely-selfspinPRD05}: the standard spin-spin interaction \cite{kidder-spineffects}, the quadrupole-monopole term arising from corrections to the Newtonian potential caused by a spinning object's mass quadrupole moment \cite{poisson-quadrupolemonopoleterm-PRD1998}, and the self-spin interaction arising from $(\text{current quadrupole})^2$ terms in the energy flux's multipole expansion \cite{gergely-selfspinPRD05}; $\beta_{2.5}$  is the 2.5PN spin-orbit term \cite{faye-buonanno-luc-higherorderspinII,*faye-buonanno-luc-higherorderspinIIerratum,*faye-buonanno-luc-higherorderspinIIerratum2}, and  $\beta_{3.0}$ is the 3PN spin-orbit tail correction \cite{faye-buonanno-luc-higherorderspinIII}.
3PN quadratic spin corrections and higher-order spin-orbit terms are not included. This analysis assumes nonprecessing spins, so all of the $\beta$ and $\sigma$ parameters are functions of $\chi_i$ and constant in time.

Eccentricity corrections to the SPA phase are included at leading order in eccentricity via an extension of the approach in \cite{krolak} to 2PN order \cite{favata-phd}:
\begin{widetext}
\begin{multline}
\Delta \Psi_{\rm 2PN}^{\rm ecc.} =
-\frac{2355}{1462} e_0^2 \left( \frac{v_0}{v} \right)^{19/3} \left[
1 + v^2 \left( \frac{299\,076\,223}{81\,976\,608} + \frac{18\,766\,963}{2\,927\,736}\eta \right)
+ v_0^2 \left( \frac{2833}{1008} - \frac{197}{36} \eta \right)
- \frac{2\,819\,123}{282\,600} \pi v^3 + \frac{377}{72} \pi v_0^3
\right. \\
+ v^4 \left( \frac{16\,237\,683\,263}{3\,330\,429\,696} + \frac{24\,133\,060\,753}{971\,375\,328} \eta + \frac{1\,562\,608\,261}{69\,383\,952} \eta^2 \right)
+ v_0^4 \left( -\frac{1\,193\,251}{3\,048\,192} - \frac{66\,317}{9072} \eta + \frac{18\,155}{1296} \eta^2  \right)
\\ \left.
+ v_0^2 v^2 \left( \frac{847\,282\,939\,759}{82\,632\,420\,864} - \frac{718\,901\,219}{368\,894\,736} \eta - \frac{3\,697\,091\,711}{105\,398\,496} \eta^2 \right)
\right].
\end{multline}
\end{widetext}
Here, $e_0$ is the eccentricity at a reference frequency $f_0=10\,{\rm Hz}$ and $v_0=(\pi M f_0)^{1/3}$.
This correction ignores periodic oscillations in the phase that occur on the orbital timescale; it also ignores harmonics of the GW signal at frequencies other than twice the orbital frequency (these are small for low eccentricity).

The tidal correction is
\be
\Delta \Psi_{\rm 6PN}^{\rm tidal} = -\frac{39}{2} \tilde{\Lambda} v^{10} + v^{12} \left( \frac{6595}{364} \delta \tilde{\Lambda} - \frac{3115}{64} \tilde{\Lambda} \right),
\ee
where the leading-order (5PN) correction depends on a ``reduced'' tidal deformability parameter $\tilde{\Lambda}$ (proportional to the $\tilde{\lambda}$ introduced in \cite{flanagan-hinderer-lovenumPRD2008}):
\begin{multline}
\label{eq:tildeLambda}
\tilde{\Lambda} \equiv 32 \frac{\tilde{\lambda}}{M^5} = \frac{8}{13} \left[ ( 1+ 7 \eta -31\eta^2) ( \hat{\lambda}_1 + \hat{\lambda}_2)
\right. \\ \left.
- \sqrt{1-4\eta} ( 1+9\eta-11\eta^2  ) (\hat{\lambda}_1 - \hat{\lambda}_2 ) \right].
\end{multline}
The relative 1PN [$O(v^{12})$] correction \cite{vines-flanagan-hinderer-1PNtidal-PRD2011} can be written in terms of $\tilde{\Lambda}$ and another parameter $\delta \tilde{\Lambda}$ \cite{[{}][{(in preparation)}]favata-tidalapproximants}. For equal-mass NSs and $\hat{\lambda}_1=\hat{\lambda}_2=\hat{\lambda}$, $\tilde{\Lambda}\rightarrow \hat{\lambda}$ and $\delta \tilde{\Lambda} \rightarrow 0$.
This ($\tilde{\Lambda}, \delta \tilde{\Lambda}$) parametrization is advantageous because the $\delta \tilde{\Lambda}$ contribution is very small ($\delta \tilde{\Lambda}/\tilde{\Lambda} \sim 0\mbox{--}0.01$) and can be ignored. This reduces the number of parameters needed and improves their measurement precision. The parameters $\hat{\lambda}_i$ span a range $\sim 60\mbox{--}1600$ for $1.4 M_{\odot}$ NSs and can reach values as large as $4400$ for $1.2 M_{\odot}$ NSs \cite{hinderer-etal-lovenum-PRD2010,lackey-shibata-etal-PRD2012-NSBH-eos}.

The test-mass limit contributions have the form $\Delta \Psi_{\rm 6PN}^{\rm tm} = \sum_{n=8}^{12} c_{n}^{\rm tm} v^n$, where $c_{n}^{\rm tm}$ are independent of $\eta$ but can be quadratic functions of $\ln v$. These coefficients are derived to 6PN order \cite{favata-tidalapproximants,varma-fujita-iyer-PRD2013} using the results of \cite{fujita-14PNEdot,*fujita-22PN,[{The 5PN and higher-order coefficients must be supplemented by the next-to-leading order corrections to the SPA derived (to leading PN order) in\,\,}][{. This adds a negligible correction $\delta c_{10} = \frac{11776}{135}\eta^2 \approx 5\times 10^{-4} c_{10}^{\rm tm} $ to the 5PN coefficient.}]poissonowen-stationaryphase}.
These coefficients $c_{n}^{\rm tm}=c_n^{\rm pp}(\eta=0)$  are taken as a proxy for the (unknown) full PN coefficients $c_n^{\rm pp}(\eta)$ for $n\geq 8$. Table \ref{tab:PNcoeffF2} justifies this approximation by showing that the $\eta \rightarrow 0$ piece of the known $c_n^{\rm pp}(\eta)$ ($n\in [2,7]$) is typically (but not always) a significant fraction of the total coefficient's value.

\begin{figure}
\includegraphics[angle=0, width=0.45\textwidth]{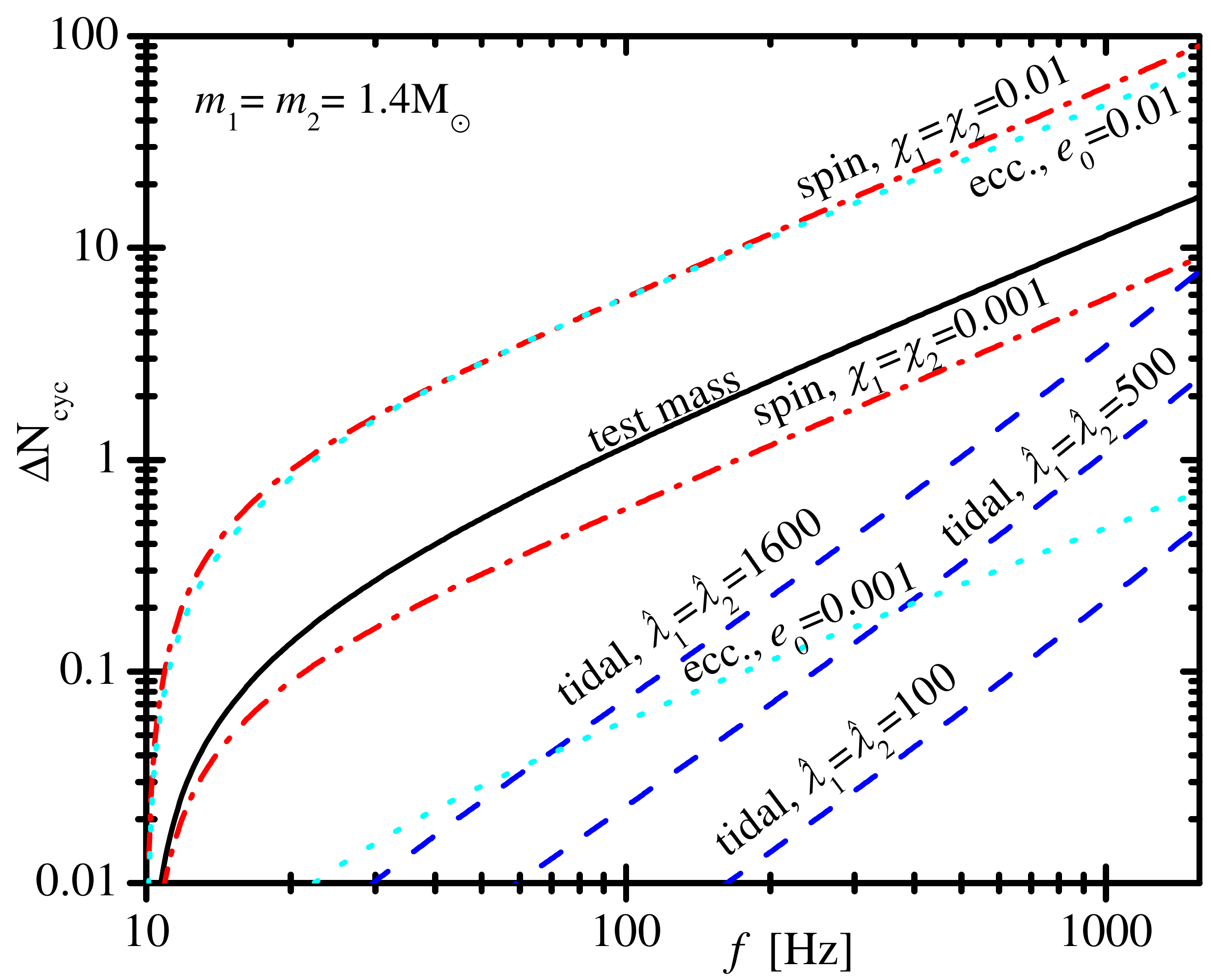}
\caption{\label{fig:fig1}(color online). Contribution of the last four terms in Eq.~\eqref{eq:Psiterms} to the number of wave cycles in the phase of the Fourier transform $\tilde{h}(f)$. Parameter values are as indicated; $t_c$ and $\phi_c$ are chosen such that $\Delta N_{\rm cyc}$ and its frequency derivative vanish at a reference frequency $f_0=10$ Hz: i.e., for each term $X$, $2\pi \Delta N_{\rm cyc}^X=\Delta \psi^X(f) - \Delta \psi^X(f_0) + (f-f_0) d\Delta \psi^X/df_0$, where $\Delta \psi^X \equiv 3 \Delta \Psi^X/(128 \eta v^5)$.}
\end{figure}

The \emph{approximate} waveform model $\tilde{h}_{\rm AP}(f) = {\mathcal A} f^{-7/6} e^{i \Psi_{\rm AP}}(f)$ has the same amplitude as $\tilde{h}_{\rm T}$ but the phase $\Psi_{\rm AP}$ is a truncation of $\Psi_{\rm T}$ as described below. Figure \ref{fig:fig1} illustrates the contribution of the last four terms in Eq.~\eqref{eq:Psiterms}, $\Delta \Psi^X = [\Delta \Psi_{\rm 6PN}^{\rm tm}, \Delta \Psi_{\rm 3PN}^{\rm spin}, \Delta \Psi_{\rm 2PN}^{\rm ecc.}, \Delta \Psi^{\rm tidal}_{\rm 6PN}]$. For plausible BNS parameter values these terms cause significant dephasing in the LIGO band; the first three terms clearly dominate the tidal term.
\begin{table}
\caption{\label{tab:PNcoeffF2}Contributions to the PN coefficients $c_{n}^{\rm pp}(\eta)$ in the SPA phase $\Delta \Psi^{\rm pp}_{\rm 3.5PN}(f)$. Columns list the PN order, the $\eta=0$ contribution to the coefficient, the equal-mass limit, and the fractional contribution of the $\eta$-dependent piece (which is $< 35\%$ except for the 3PN piece).
Logarithmic terms are evaluated at $v=0.351$ ($f=1000\, {\rm Hz}$ for two $1.4 M_{\odot}$ NSs).}
\begin{tabular}{|l|r|r|r|}
\hline \hline
PN coeff. & $c_{n}^{\rm pp}(0)$ & $c_{n}^{\rm pp}(0.25)$ & $1-\frac{c_{n}^{\rm pp}(0)}{c_{n}^{\rm pp}(0.25)}$ \\
\hline
1PN & $4.91$ & $6.44$  & $0.237$ \\
1.5PN & $-50.3$  &  $-50.3$  & $0$ \\
2PN & $30.1$ &  $46.2$ & $0.349$ \\
2.5PN & $-504$  & $-487$  & $-0.0366$ \\
3PN & $63.2$  & $-763$  & $1.08$ \\
3.5PN & $954$ & $1131$  & $0.157$ \\
\hline \hline
\end{tabular}
\end{table}

\emph{Statistical and systematic errors}.---To compute statistical errors we use the Fisher matrix formalism \cite{flanagancutler,*poisson-will-2PNparameterestimate,[{For low mass binaries ($M \lesssim 4 M_{\odot}$) with ${\rm SNR}\gtrsim 10$, Fisher matrix error estimates of $M_{\rm ch}$ and $\eta$ are comparable to MCMC calculations; c.f.~Fig.~1 and Table III of }][{}]rodriguez-farr-farr-mandel-fisherinadequacies}.
The Fisher matrix is $\Gamma_{ab} \equiv \left( \partial_a h_{\rm AP}| \partial_b h_{\rm AP} \right)$, where $\partial_a \equiv \partial/\partial \theta_a$ and the inner product weighted by the detector noise spectral density $S_n(f)$ \cite{[{I use a fit to the zero-detuned high-power aLIGO noise from Eq.~(4.7) of }][{}]ajith-spin-PRD2011,*[{the ET sensitivity is from Table I of }][{}]sathya-schutz-LLR2009} is $(a|b) \equiv 2 \int_{f_{\rm low}}^{f_{\rm high}} df S_n^{-1} [\tilde{a} \tilde{b}^{\ast} + \tilde{a}^{\ast} \tilde{b}]$ ($\ast$ denotes complex conjugation). The statistical error on the parameter $\theta_a$ is given by $\delta \theta_a = (\Gamma^{-1}_{aa})^{1/2}$. (Gaussian priors on $\theta_a$ are incorporated by adding a component $\Gamma_{aa}^0=\sigma_a^{-2}$ to $\Gamma_{ab}$ \cite{flanagancutler,*poisson-will-2PNparameterestimate}, where $\sigma_a$ is the maximum bound on $\delta \theta_a$.) The formalism of \cite{cutler-vallisneri-systematicerrors-PRD2007} is used to compute the systematic error. This involves minimizing the inner product $\left(h_{\rm T}(\theta^{\rm T}_a) - h_{\rm AP}(\theta_a^{\rm best\,fit}) | h_{\rm T}(\theta^{\rm T}_a) - h_{\rm AP}(\theta_a^{\rm best\,fit}) \right)$, which represents the distance in parameter space between the approximate waveform (evaluated at the best-fit parameter values) and the true waveform (evaluated at the true parameter values). From Eq.~(29) of \cite{cutler-vallisneri-systematicerrors-PRD2007} the systematic error can be approximated as
\begin{align}
\Delta \theta_a &\approx \Gamma^{-1}_{ab} \left(\partial_b h_{\rm AP} | h_{\rm T} - h_{\rm AP} \right),\\
&=4 {\mathcal A}^2 \Gamma^{-1}_{ab} \int_{f_{\rm low}}^{f_{\rm high}} df\, \frac{f^{-7/3}}{S_n(f)} (\Psi_{\rm T} - \Psi_{\rm AP}) \partial_b \Psi_{\rm AP},
\end{align}
where the second line follows from our SPA model.

Statistical and systematic parameter errors were computed for a $(1.4+1.4) M_{\odot}$ BNS at $100$ Mpc, assuming a single Advanced LIGO (aLIGO) or Einstein Telescope (ET) interferometer.
The parameter set is $\theta_a=[t_c, \phi_c, \ln M_{\rm ch}, \ln\eta]$ ($M_{\rm ch}=\eta^{3/5} M$ is the chirp mass), with priors $\sigma_{\eta}=0.25$ and $\sigma_{\phi_c}=\pi$ and  integrating from $f_{\rm low}=10 \,{\rm Hz}$ (aLIGO) or $1 \,{\rm Hz}$ (ET) to $f_{\rm high}=1000 \,{\rm Hz}$.

The following five scenarios were considered:

(i) To study the effect of varying the PN order we choose $\Psi_{\rm T} \propto \Delta \Psi_{\rm 3.5PN}^{\rm pp} + \Delta \Psi_{\rm 6PN}^{\rm tm}$. The same expression for $\Psi_{\rm AP}$ is used except the PN series is sequentially truncated at increasing powers of $v$ (from 1PN to 5.5PN).

(ii) Neglecting NS spin is modeled by choosing $\Psi_{\rm T} \propto \Delta \Psi_{\rm 3.5PN}^{\rm pp} + \Delta \Psi_{\rm 3PN}^{\rm spin}$ and $\Psi_{\rm AP} \propto \Delta \Psi_{\rm 3.5PN}^{\rm pp}$ while varying $\chi_i$ (assumed to be equal for each NS).

(iii) To neglect eccentricity we choose $\Psi_{\rm T} \propto \Delta \Psi_{\rm 3.5PN}^{\rm pp} +\Delta \Psi_{\rm 2PN}^{\rm ecc.}$ and $\Psi_{\rm AP} \propto \Delta \Psi_{\rm 3.5PN}^{\rm pp}$, varying $e_0$.

(iv) To neglect tidal interactions we choose $\Psi_{\rm T} \propto \Delta \Psi_{\rm 3.5PN}^{\rm pp} + \Delta \Psi_{\rm 6PN}^{\rm tidal}$ and $\Psi_{\rm AP} \propto \Delta \Psi_{\rm 3.5PN}^{\rm pp}$ while varying $\hat{\lambda}_i$ (also set equal for each NS).

(v) Last, we consider the neglect of spin, eccentricity, or high-PN-order terms on the tidal deformation measurability. We choose $\Psi_{\rm T} \propto \Delta \Psi_{\rm 3.5PN}^{\rm pp} + \Delta \Psi_{\rm 6PN}^{\rm tidal} + \Delta \Psi^X$ and $\Psi_{\rm AP} \propto \Delta \Psi_{\rm 3.5PN}^{\rm pp} + \Delta \Psi_{\rm 6PN}^{\rm tidal}$ while varying $\hat{\lambda}_i$. Here, $X$ refers to the spin, ecc., or tm terms in Eq.~\eqref{eq:Psiterms}. We set $\delta \tilde{\Lambda}=0$, and $\tilde{\Lambda}$ is added as a parameter to $\Gamma_{ab}$ with prior $\sigma_{\tilde{\Lambda}}=5000$.

\begin{SCfigure*}
$
\begin{array}{cc}
\includegraphics[angle=0, width=0.35\textwidth]{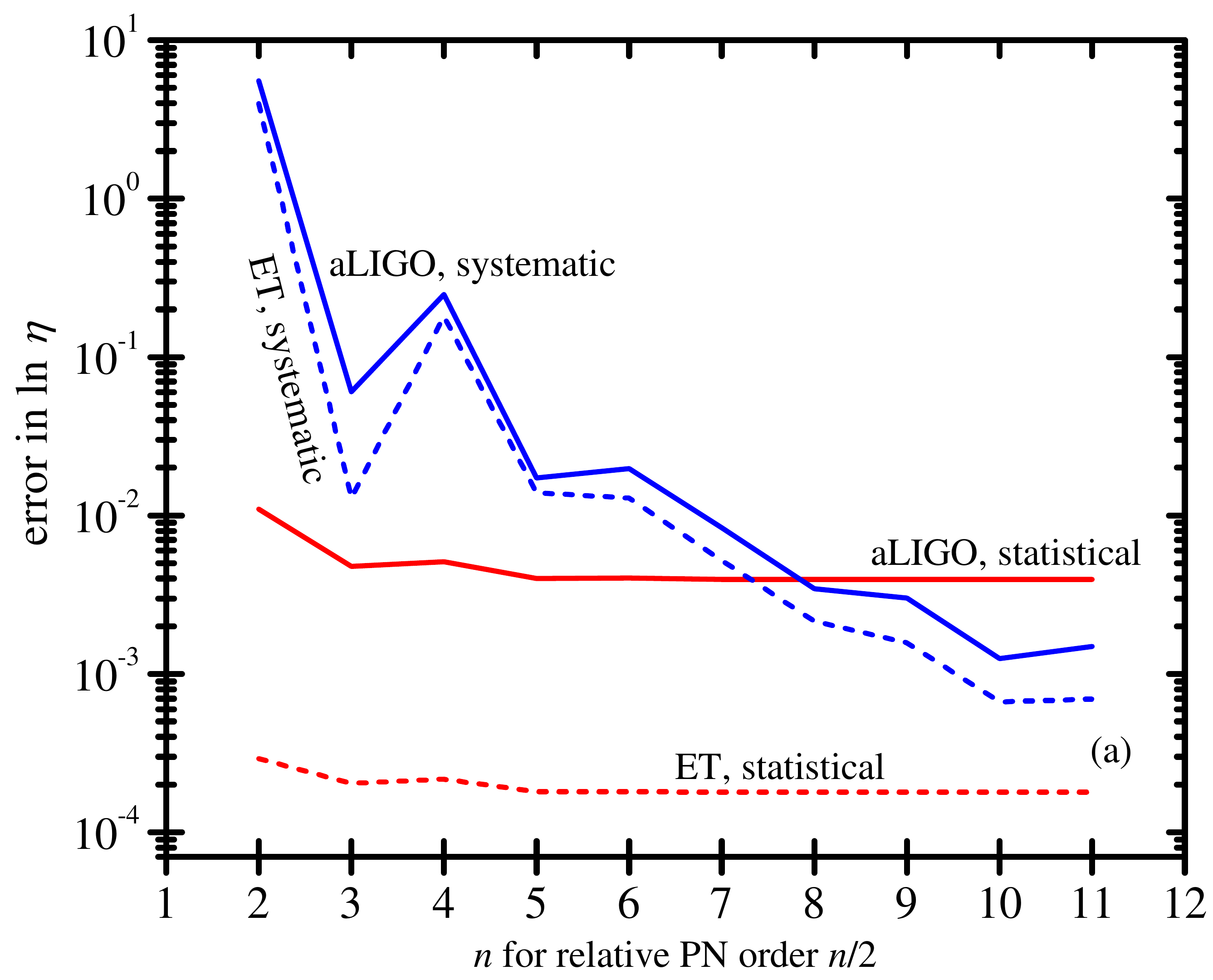} &
\includegraphics[angle=0, width=0.35\textwidth]{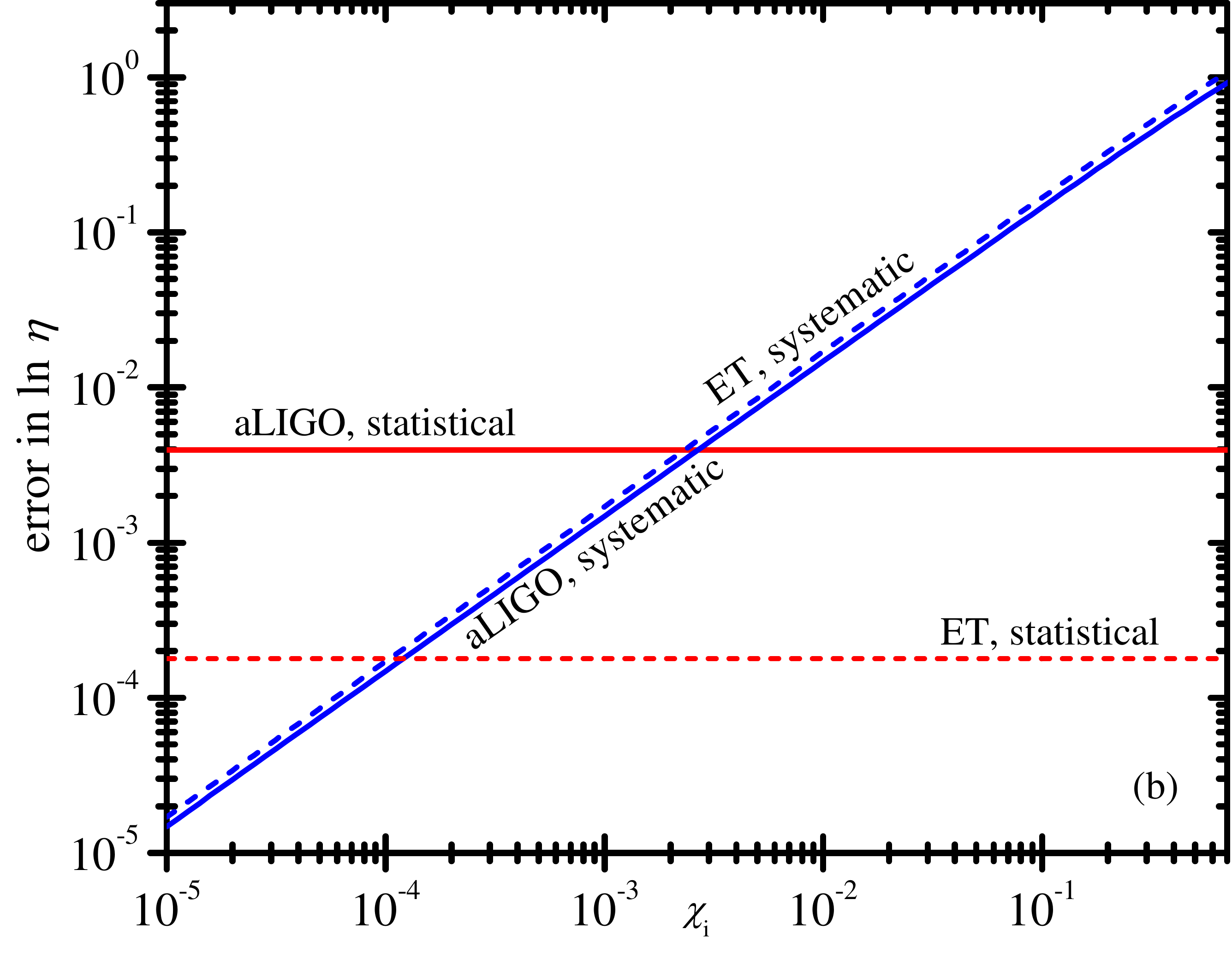} \\
\includegraphics[angle=0, width=0.36\textwidth]{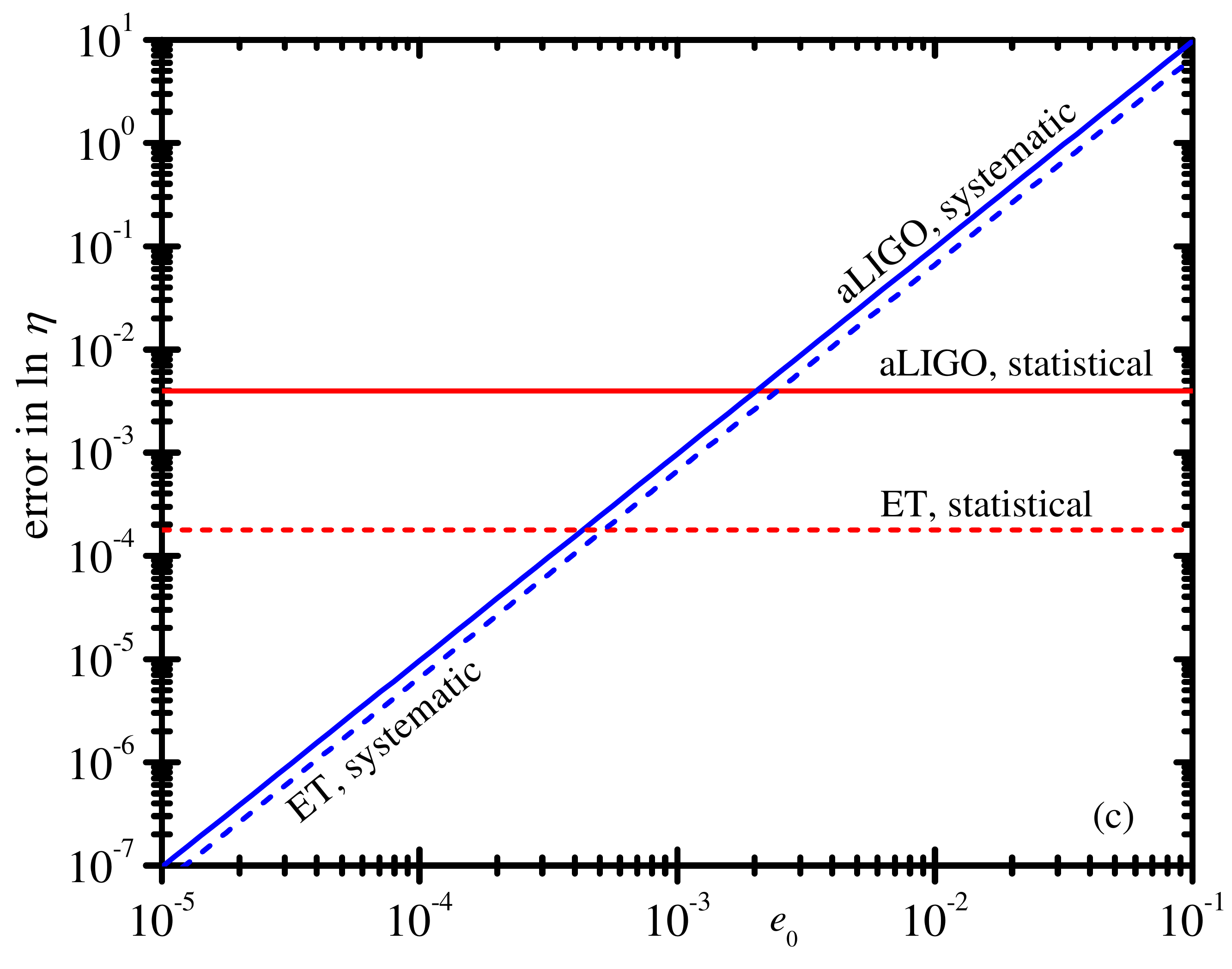} &
\includegraphics[angle=0, width=0.35\textwidth]{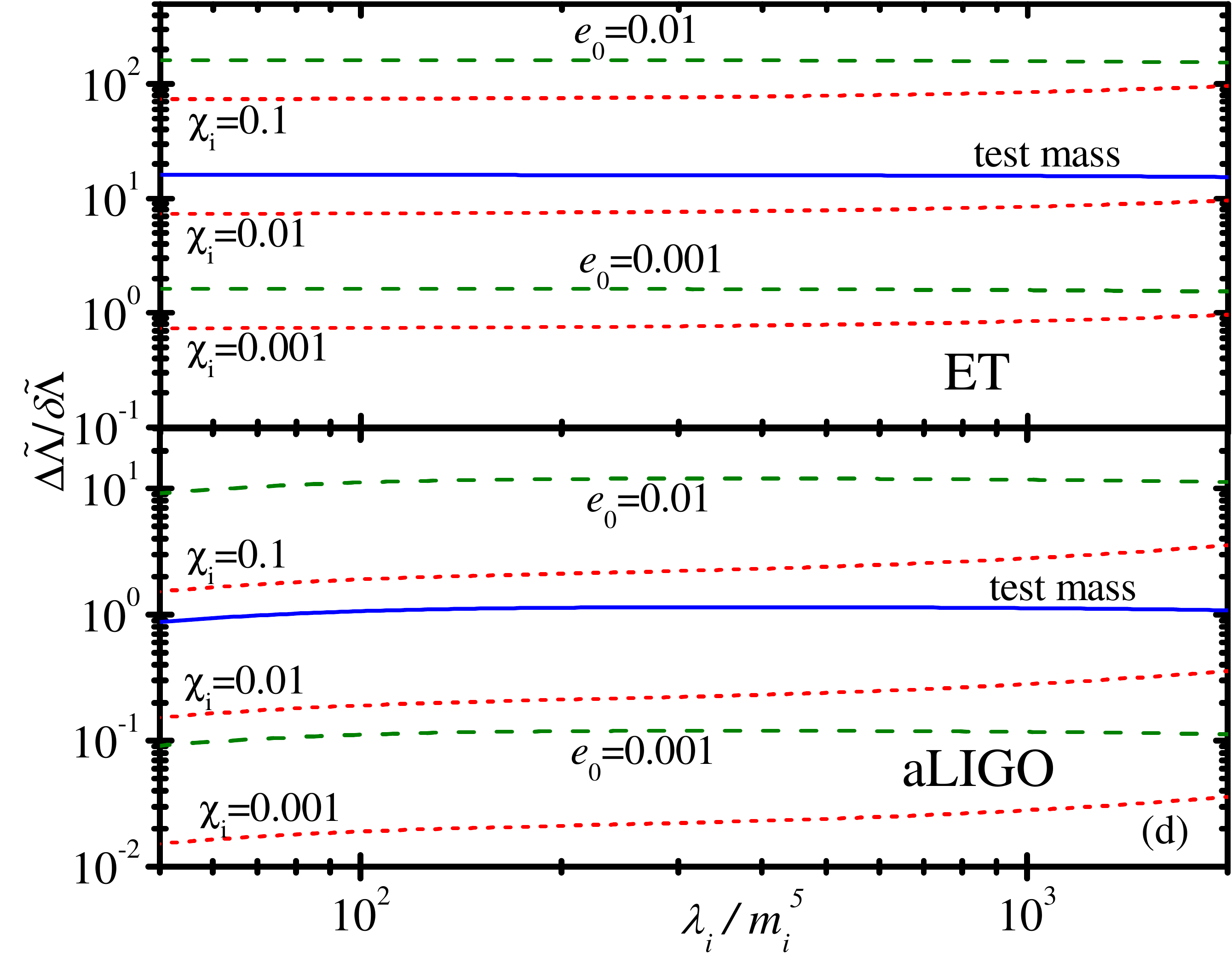}
\end{array}
$
\centering
\vspace{-0.7cm}
\caption{\label{fig:PEresults}(color online). aLIGO and ET statistical and systematic parameter errors for different waveform approximations. (a) Fractional errors in $\eta$ due to neglecting high PN-order terms as a function of the relative PN order in $\Psi_{\rm AP}$. (b) The corresponding error due to neglecting spin effects. (c) The error from neglecting orbital eccentricity. (d) The ratio of systematic to statistical errors in the reduced tidal deformation parameter $\tilde{\Lambda}$ as a function of $\hat{\lambda}_i$ and for different waveform errors (neglecting spin, eccentricity, or high PN terms). All plots are for a $m_1=m_2 = 1.4 M_{\odot}$ BNS at $100$ Mpc. Angle-averaged SNRs are $14.5$ for aLIGO and 175 for ET. These results are relatively insensitive to small changes in $\eta$ that keep $M$ constant.}
\end{SCfigure*}
\emph{Results}.---Figure \ref{fig:PEresults} summarizes the most important results of this study, focusing on the fractional errors in the reduced mass ratio.
Figure \ref{fig:PEresults}(a) shows the statistical and systematic errors for case (i). Here we see that using an approximate 3.5PN template results in a systematic error that is double the statistical error, $\Delta \eta / \delta \eta \approx 2.1$ (aLIGO). Figure \ref{fig:PEresults}(a) suggests that decreasing the systematic error on the masses below the statistical errors will require 4PN order waveforms. For ET, waveforms will likely need to be known to at least 6PN order. Figure \ref{fig:PEresults}(b) shows error estimates for case (ii) as a function of the NS spin parameter. Here we see that dimensionless spins as small as $\chi_i \approx 0.003 \, (10^{-4})$ can cause systematic errors to exceed statistical errors for aLIGO (ET). Figure \ref{fig:PEresults}(c) shows error estimates for case (iii) as a function of the orbital eccentricity at $10 \,{\rm Hz}$. Here we see that neglecting eccentricity causes systematic errors that exceed statistical errors if $e_0 \gtrsim 0.002 \, (0.0005)$ for aLIGO (ET). While systematic errors in $M_{\rm ch}$ can also exceed statistical ones, in nearly all cases both errors are $\ll 1\%$.

For case (iv) $\Delta \eta / \eta < 3\%$ but exceeds $\delta \eta / \eta$ for $\hat{\lambda}_i>320$ \cite{[{Note that }][{  previously showed that tidal effects will not affect \emph{detection}.}]pannarale-etal-PRD2011-BHNS-eos}. For case (v) [Fig.~\ref{fig:PEresults}(d)] we see that the unknown 4PN and higher point-particle terms nearly always introduce a systematic bias in $\tilde{\Lambda}$ [$\Delta \tilde{\Lambda} / \delta \tilde{\Lambda} \sim 1.1 \, (16)$ for aLIGO (ET)], making that parameter unmeasurable with existing PN waveforms. This plot also indicates that spins $\chi_i \gtrsim 0.03$ or eccentricities $e_0 \gtrsim 0.003$ yield $\Delta \tilde{\Lambda} > \delta \tilde{\Lambda}$ (aLIGO). This poses difficulties for attempts to infer information about the distance-redshift relation from BNS inspirals \cite{messenger-read-PRL2012}. Note that the statistical errors $\delta \tilde{\Lambda} / \tilde{\Lambda} \sim 20\% \mbox{--} 600\%$ for aLIGO are themselves large.

Statistical errors for ET are smaller by a factor $\sim 10$ (largely due to higher SNR). Because systematic errors are SNR independent, understanding waveform errors is especially critical for third generation detectors like ET.

\emph{Conclusions}.---Several plausible theoretical waveform errors were considered, and the corresponding systematic biases were evaluated and found to be important in many cases. Correcting these biases will be essential for accurate parameter extraction from expected aLIGO observations. In particular the following suggestions are recommended: (1) standard PN waveforms should be developed to 4PN order to improve mass accuracy, and to 5PN order to reduce biases on equation-of-state constraints; (2) while their dimensionless spins are much smaller than for black holes, NS spin effects should not be neglected in parameter estimation studies; (3) although eccentricities of known binary pulsars will be smaller than $e_0 \lesssim 10^{-5}$ at 10 Hz, eccentric templates will be needed if astrophysical scenarios predict $e_0\gtrsim 10^{-3}$. Further investigations of these issues are in progress \cite{favata-tidalapproximants}.

This work was supported by NSF Grants No.~PHY-0970074, No.~PHY-1308527, and the UWM Research Growth Initiative. I thank Curt Cutler, Stephen Fairhurst, Jocelyn Read, Michele Vallisneri, and the referees for comments on this manuscript. I also thank Kent Yagi and Nicol\'{a}s Yunes for helping me find a coding error that affected my results. Their complementary work \cite{kent-nico-systematictidal} focuses on the systematic error in the tidal deformability using a modified version of the scheme discussed here.

\bibliography{systematicerror-textV2b}
\end{document}